\documentclass[10pt,letterpaper,twocolumn]{article} 

\usepackage{ol}
\usepackage{graphicx}
\usepackage{amsmath}

\begin{document}

\twocolumn[ 

\title{Polarization singularity anisotropy: determining monstardom}

\author{Mark R Dennis}

\address{H H Wills Physics Laboratory, University of Bristol, Tyndall Avenue, Bristol BS8 1TL, UK}

\begin{abstract}
C points, that is isolated points of circular polarization in transverse fields of varying polarization, are classified morphologically into three distinct types, known as lemons, stars and monstars.
These morphologies are interpreted here according to two natural parameters associated with the singularity, namely the anisotropy of the C point, and the polarization azimuth on the anisotropy axis.
In addition to providing insight into singularity morphology, this observation applies to the densities of the various morphologies in isotropic random polarization speckle fields.
\end{abstract} 


]

Singular optics is the study of topological features of optical fields, and their organization of the local and global field morphology.
In 2-dimensional transverse fields, the features usually emphasized are points, such as optical vortices (nodes, phase singularities) in scalar fields [\onlinecite{nb:1974b34,nye:1999natural}], and points of circular polarization -- C points -- in elliptically polarized vector fields [\onlinecite{nye:1983cline}].
These singularities possess a topological index: a quantized, positive or negative (half-)integer, for the number of phase or polarization azimuth cycles around the singularity.
The field surrounding a singular point has a typical geometric structure: an elliptic core of darkness for vortices, and various morphologies of C point: lemon and monstar (positive index), and star (negative index).

This latter distinction was considered in Ref.~\onlinecite{bh:1977b60}, following work of Darboux [\onlinecite{darboux:1896lecons4}], in the geometry of umbilic points on surfaces (also see, for instance, Ref.~\onlinecite{porteous:geometric}).
There, the points are singularities of directions of principal curvature (axes of the Gauss curvature ellipse [\onlinecite{porteous:geometric}]), whose pattern is very similar to the pattern of polarization ellipse azimuths.
The condition distinguishing lemons from monstars [\onlinecite{bh:1977b60}] was translated into polarization language [\onlinecite{dennis:2002polarization}] as a complicated expression (Eq.~(\ref{eq:dl}) below).
In this Letter I return to this definition.
By considering C points as vortices in a circular component of the polarization, I will show that monstars may be interpreted purely as particularly anisotropic positive index C points.

In a general elliptically polarized, coherent 2-dimensional field, the position-dependent state of polarization is conveniently represented by
\begin{equation}
   \sigma \equiv S_1 + i S_2 = E_{\mathrm{R}}^* E_{\mathrm{L}}.
   \label{eq:stokes}
\end{equation}
where $S_1$ and $S_2$ are the two Stokes parameters independent of the sense of circular polarization, and $E_{\mathrm{R}}$ ($E_{\mathrm{L}}$) is the right- (left-)handed circular component at that point.
$|\sigma|$ is proportional to the polarization ellipticity, and $\arg\sigma$ is twice the polarization azimuth $\theta$ (angle between the polarization ellipse major axis and the $x$-axis).
This parametrization goes back to Poincar{\'e} [\onlinecite{poincare:1892lumiere}], and is often used in polarization singularity physics [\onlinecite{dennis:2002polarization,km:2001analogy,fsm:2002elliptic}].

At a C point, $\sigma = 0$ and the azimuth is undefined.
At such a point, either $E_{\mathrm{R}}$ or $E_{\mathrm{L}}$ is zero (not both), i.e.~there is a vortex in that component.
It will be assumed that $E_{\mathrm{L}} = 0$ (a right-handed C point), with similar arguments holding in the other case.

\begin{figure}
   \centerline{\includegraphics[width=7.2cm]{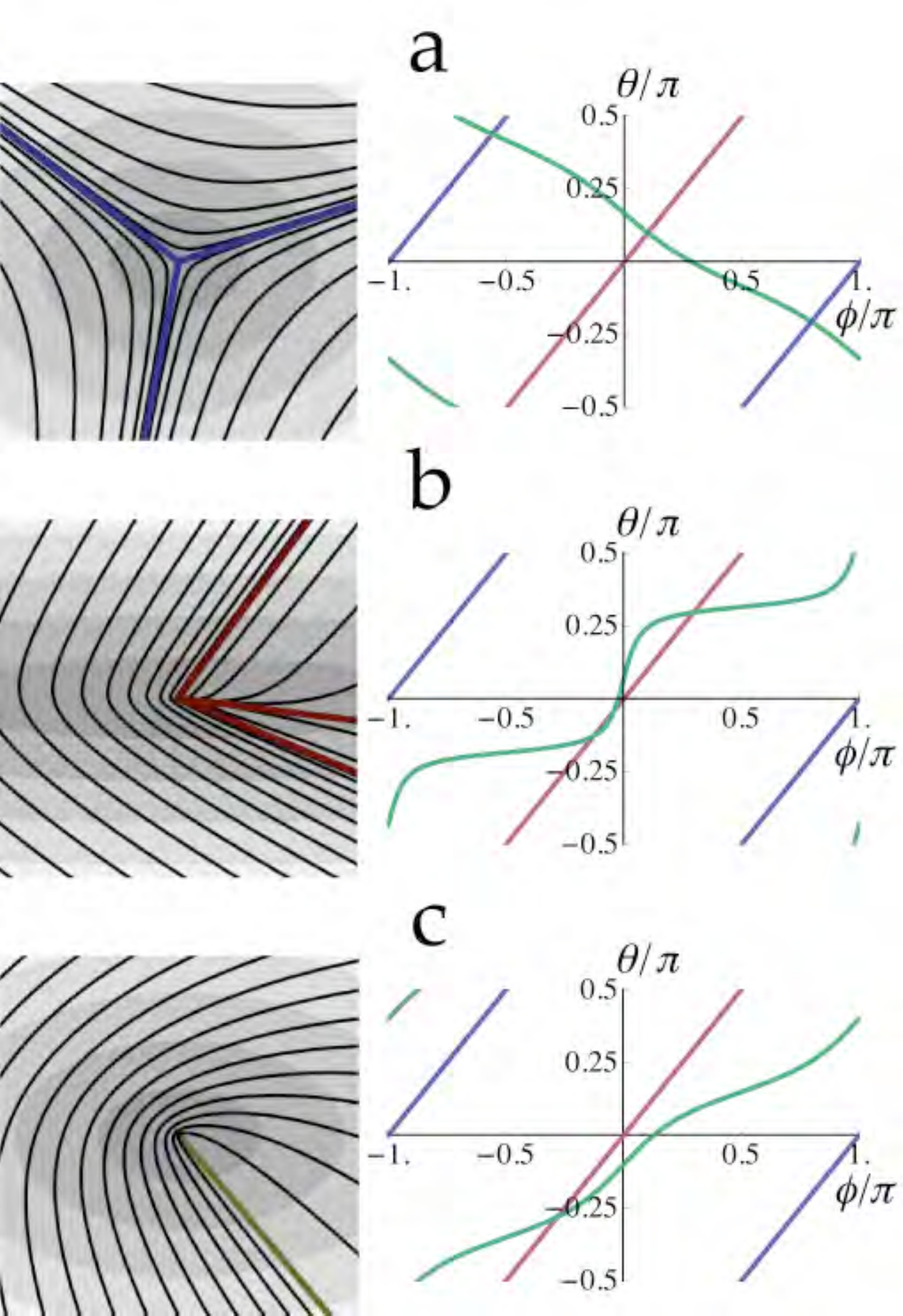}}
   \caption{(Color online) Figures showing the three different line types of C point morphology with typical shape parameters, and corresponding plots of $\theta(\phi):$ (a) star ($\Upsilon = -.95, \beta = \pi/6$); (b) monstar ($\Upsilon = .31, \beta = \pi/15$); (c) lemon ($\Upsilon = .87, \beta = -\pi/10$).
   The plots on the left-hand sign show grayscale contour plots of $|\sigma|,$ showing the elliptic anisotropy, and the polarization lines (black) with colored separatrices.
   The straight lines in the plots on the right-hand side correspond to the lines $\theta = \phi$ (purple), $\phi \pm \pi$ (blue), which intersect the curve $\theta(\phi)$ (green) at the straight line separatrices ending at the singularity.
   $\theta(\phi),$ being periodic, multiply wraps around the plot.
   }
   \label{fig:lms}
\end{figure}

The azimuth structure of an elliptically polarized field can be made clear by plotting the `polarization lines' -- families of curves whose tangent gives the polarization azimuth.
Although the lines can be found analytically in some cases [\onlinecite{bdl:2004b373,bd:2003b355}], they usually have to be calculated numerically.
Polarization lines terminate on C points, either in threes (a {\em star}, Fig.~\ref{fig:lms}(a)), infinitely many, with three straight (a {\em monstar}, Fig.~\ref{fig:lms}(b)), or one only (a {\em lemon}, Fig.~\ref{fig:lms}(c)).
Mathematically, these straight lines terminating on the singularity are separatrices, separating regions of polarization lines with differently-signed curvature.
The line classification (darbouxian, or L classification), that is three separatrices (star, monstar) or one (lemon) is given by the sign of the quantity $D_L,$ with $x,y$ subscripts denoting partial derivatives:
\begin{multline}
    D_L = \left[(2S_{1y} + S_{2x})^2 - 3 S_{2y}(2S_{1x} - S_{2y})\right]
    \\ \times \left[(2S_{1x} - S_{2y})^2 + 3 S_{2x}(2S_{1y} + S_{2x})\right]\\
     -(2S_{1x} S_{1y} + S_{1x} S_{2x} - S_{1y} S_{2y} + 4S_{2x} S_{2y})^2
    \label{eq:dl}
\end{multline}
(positive for three, negative for one) [\onlinecite{dennis:2002polarization}]. 
Around the C point, the polarization azimuth turns through $\pm \pi$ (i.e.~index $\pm 1/2$): $+$ for lemon and monstar, $-$ for star.
Monstars therefore have a transitional nature, with the same index as a lemon and the same line classification as a star, hence its name (le)monstar [\onlinecite{bh:1977b60}].

The index of the C point is determined by the sign of the following quantity:
\begin{equation}
   \Upsilon \equiv \frac{\operatorname{Im} \nabla\sigma^{\ast}\times \nabla\sigma}{|\nabla\sigma|^2} = \frac{2(S_{1x} S_{2y} - S_{2x} S_{1y})}{S_{1x}^2 + S_{1y}^2 + S_{2x}^2 + S_{2y}^2}.
   \label{eq:isotropy}
\end{equation}
The middle expression is the usual expression for the vortex sign in $\sigma$ (it is also the sign of the vortex in $E_{\mathrm{L}}$).
The normalization gives $\Upsilon$ as a dimensionless, signed measure of the ellipticity (isotropy) of the contours of $|\sigma|$ around the singularity: $\pm 1$ for isotropic, approaching 0 for extreme anisotropy.
Stokes-like parameters in the derivatives of complex scalar fields have previously been used to describe the anisotropy of optical vortices [\onlinecite{dennis:twirl,efv:fine,roux:coupling}].
Viewing the C point as a nodal point of $\sigma,$ $\Upsilon$ is analogous to the normalized, third Stokes parameter (which determines the polarization ellipse eccentricity).

The anisotropy of the singularity also determines the dependence of the azimuth $\theta$ in terms of the polar angle $\phi$ around the C point.
It is simple to show that
\begin{equation}
   \theta(\phi) = \frac{1}{2} \arctan \frac{S_{2x} \cos\phi + S_{2y} \sin\phi}{S_{1x} \cos\phi + S_{1y} \sin\phi}.
   \label{eq:thetaphi}
\end{equation}
Fig.~\ref{fig:lms} depicts anisotropic C points of the different morphological types, along with elliptic $|\sigma|$ contours and the corresponding plot of $\theta$ against $\phi.$
The line classification is determined by the number of intersections of the curve $\theta(\phi)$ with the lines $\phi, \phi\pm \pi.$
The curve for the star, where $\theta(\phi)$ is monotonic decreasing, always intersects three times.
The positive index points (monotonic increasing) have one or three intersections, depending on the shape of the curve (detemined by $\Upsilon$) and the intercept $\theta(\phi = 0).$
On increasing the intercept $\theta(0)$ in (b), two adjacent intersections vanish and the monstar becomes a lemon [\onlinecite{nye:1999natural}].
The curve in (c) is too straight to give three intercepts, whatever its intercept.

This relationship may be clarified by choosing coordinates $X, Y$ based on the axes of the anisotropy ellipse of $|\sigma|$ contours, with $X$ ($Y$) the major (minor) axis, where
\begin{equation}
   \nabla \sigma = \exp(2 i \beta) \{T_X, i T_Y\}
   \label{eq:Tfix}
\end{equation}
with $|T_Y| \ge T_X \ge 0.$
$\beta$ is the angle between the polarization azimuth angle (with respect to the $X$-axis) and the $X$ axis, and $-\pi/4 \le \beta = \tfrac{1}{2}\arctan S_{2X}/S_{1X} \le \pi/4.$
This is the coordinate system in Fig.~\ref{fig:lms}, so $\beta = \theta(\phi = 0)$ (with $\phi = 0$ on the +$X$-axis).
$\Upsilon,$ as a shape measure, is unaffected by this choice.
The shape of the curve $\theta(\phi)$ is determined completely by $\Upsilon,$ and, in $X,Y$ coordinates,
\begin{equation}
   \theta(\phi) = \frac{1}{2} \arctan\left[ \frac{\Upsilon \sin\phi}{(1 - \sqrt{1-\Upsilon^2}))\cos \phi}\right] + \beta.
   \label{eq:thetaXY}
\end{equation}

\begin{figure}
   \centerline{\includegraphics[width=9cm]{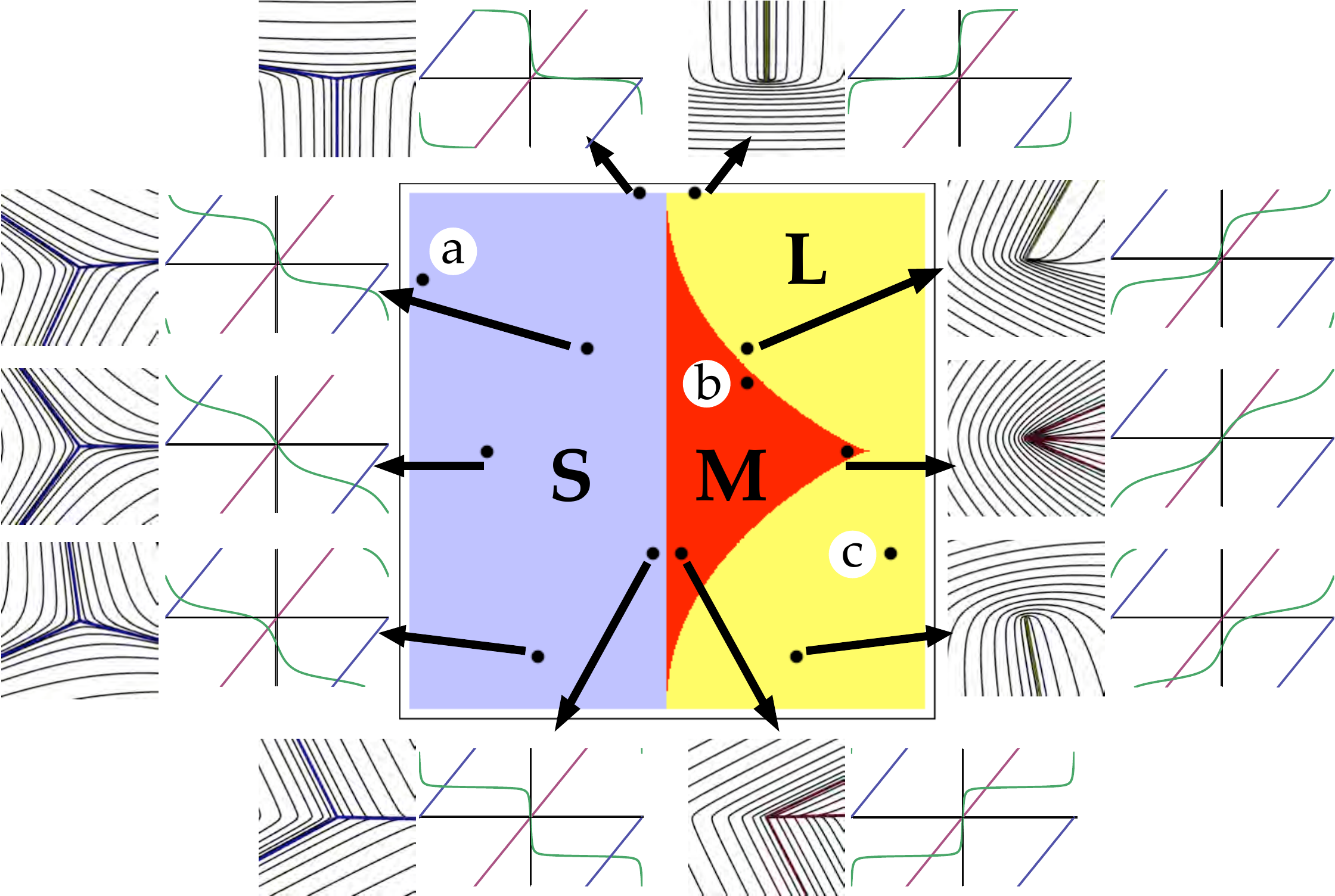}}
   \caption{(Color online) The $\Upsilon,\beta$ plane parametrizing C point morphology ($-1 \le \Upsilon \le 1, -\pi/4 \le \beta \le \pi/4$), together with several choices of singularity parameters, plotted as Fig.~\ref{fig:lms}.
   The plane is partitioned into three distinct regions corresponding to star (S, blue), monstar (M, red) and lemon (L, yellow).
   C point geometries at different points of this area are also shown.
   The points labeled (a), (b) and (c) correspond to those parts of Fig.\ref{fig:lms}, and the other choices, from maximum to minimum $\beta,$ have coordinates $(\Upsilon, \beta): (\pm .10,\pi/4), (\pm .31,\pi/10), (\pm .05, -\pi/10), (\pm .5,-\pi/5).$
   }
   \label{fig:plane}
\end{figure}

The distinction between lemon and monstar is completely determined by $\Upsilon$ and $\beta.$
From Eqs.~(\ref{eq:isotropy}) and (\ref{eq:Tfix}), $D_L$ in Eq.~(\ref{eq:dl}) may be rearranged:
\begin{equation}
    D_L = \tfrac{3}{4}|\nabla \sigma|^4  (2 + \Upsilon (11 \Upsilon - 14) + 2 (1 + \Upsilon) \sqrt{1 - \Upsilon^2} \cos4\beta).
    \label{eq:Dfixed}
\end{equation}
This equation is one of the main results of this Letter.
It shows that, for given $\beta$ there is a minimum isotropy $\Upsilon_{\mathrm{crit}}$ beyond which a lemon becomes a monstar; the largest $\Upsilon_{\mathrm{crit}}$ is 4/5, occurring for $\beta = 0.$ 
Equivalently, for fixed $\Upsilon,$ there is a critical $\beta_{\mathrm{crit}},$
\begin{equation}
   \cos4\beta_{\mathrm{crit}} = - \frac{2+\Upsilon(11\Upsilon - 14)}{2(1+\Upsilon)^{3/2} (1-\Upsilon)^{1/2}},
   \label{eq:betacrit}
\end{equation}
which determines the lemon/monstar nature of a positive index C point.
This line classification in the $\Upsilon, \beta$ plane, with several C point polarization line examples for different choices of $\Upsilon, \beta$ is illustrated in Fig.~\ref{fig:plane}.
The boundary line between the monstar and lemon areas of the $\Upsilon, \beta$ plane is a cusp catastrophe: pairs of neighboring separatrices approach and annihilate as either fold line is crossed [\onlinecite{nye:1999natural}], and the three separatrices coalesce at the cusp point itself.

Further analysis in the $\Upsilon,\beta$ plane, or of the $\theta(\phi)$ curves from Eq.~(\ref{eq:thetaXY}), determines the angles of the polarization lines for monstars and stars; some examples are given in Fig.~\ref{fig:plane}.
For instance, it is straightforward to show that the maximum monstar opening angle is $90^{\circ},$ and the greatest angle between two star lines is $180^{\circ},$ when $\Upsilon \to 0$ (independent of $\beta$).
In fact, the observation that C point annihilation occurs only between monstars and stars [\onlinecite{nye:1999natural}] (i.e.~a lemon becomes a monstar as it approaches a star) can be understood as follows: as oppositely signed singularity points approach, they become more anisotropic, with $\Upsilon \to 0$ as they annihilate; for almost all $\beta,$ the $\Upsilon >0$ neighborhood of the $\Upsilon = 0$ line is of monstar type, apart from the nongeneric case $\beta = \pm \pi/4.$

Random polarization fields are a natural physical situation in which C points arise naturally, and there has been interest in the distribution of different types of C point in this case [\onlinecite{dennis:2002polarization,fsm:2002elliptic,fodp:2008polarization,des:2004measurement,sde:2004topological,vs:2008topological}].
The density fractions of lemons, stars and monstars for statistically isotropic vector speckle can easily be derived from the observations above.
As a scalar vortex ellipticity [\onlinecite{bd:2000b321}], $\Upsilon^2$ is uniformly distributed between 0 and 1 (and $\Upsilon$ is equally distributed in sign).
$\beta$ is also uniformly distributed, and is independent of $\Upsilon.$ 
Thus the fraction of stars is $1/2,$ and, from Eq.~(\ref{eq:betacrit}), the fraction of lemons is $1/\sqrt{5},$ agreeing with the numerical value of $0.447$ previously calculated [\onlinecite{bh:1977b60,dennis:2002polarization}].

These observations apply to any half-integer index singularity of a director (unoriented line segment) in two dimensions, such as the umbilic points mentioned above, and disclinations in nematic liquid crystals.
Monstar-type disclinations are not seen, probably because energetics favor defects with isotropic neighborhoods.

This parametrization may be generalised to defects in fields with an order parameter with any degree $s$ of rotational symmetry, such as fourfold-symmetric fields of crosses [\onlinecite{nelson:defects}] (occurring optically as singular axes of biaxial, dichroic crystals [\onlinecite{bd:2003b355}]).
The defect index is $\pm 1/s,$ with lemon analogs having $s-1$ separatrices, star/monstar analogs having $s+1.$
With generalized Stokes parameters, $\Upsilon$ and $\beta$ can be defined, and $\theta(\phi)$ is given by Eqs.~(\ref{eq:thetaphi}), (\ref{eq:thetaXY}) with the $1/2$ prefactor replaced by $1/s.$ 
However, the generalized $D_L$ is more complicated than Eqs.~(\ref{eq:dl}), (\ref{eq:Dfixed}) (although it depends only on $\Upsilon$ and $\beta$).

There is another classification of C points, namely the catastrophe (or contour) classification, which describes the orientation of the double-cone diabolo of the magnitudes of the polarization ellipse principal axes near the singularity [\onlinecite{nye:1999natural,nye:1983cline,bh:1977b60,dennis:2002polarization,fodp:2008polarization,vs:2008topological,esf:2006diabolos}]. 
This classification involves more parameters than the darbouxian line classification described here; in addition to $\nabla \sigma,$ it depends on the gradient of overall field intensity [\onlinecite{dennis:2002polarization}], and cannot be described directly in terms of the $\Upsilon,\beta$ plane.
Nevertheless, the parametrization presented here, through which the rather opaque line classification is interpreted simply in terms of the anisotropy of the C point and one polarization azimuth value around the singularity, should aid the geometric interpretation of this important aspect of the interpretation of polarization patterns.


I am grateful to Florian Flossmann, John Nye, Marat Soskin, Vincenzo Vitelli and Wei Wang for discussions.
My research is supported by the Royal Society of London.

\end{document}